\documentclass[sigconf]{acmart}
\usepackage{microtype}
\usepackage{comment}
\usepackage{subcaption}
\usepackage{graphicx}
\graphicspath{{./figures/}}
\usepackage{amsmath}

\usepackage{booktabs} 

\definecolor{gray97}{gray}{.97}
\definecolor{gray75}{gray}{.75}
\definecolor{gray45}{gray}{.45}

\usepackage{listings}

\lstdefinestyle{C}
   {language=C++,
   }

\begin{document}
%
\title{Parallelizing Workload Execution in Embedded and High-Performance Heterogeneous Systems}



\acmConference[HIP3ES 2018]{6th Workshop on High Performance Energy Efficient Embedded Systems}{January 2018}{Manchester UK} 
\acmYear{2018}
\copyrightyear{2018}


\author{Jose Nunez-Yanez,\\ Mohammad Hosseinabady, \\ Moslem Amiri}
\affiliation{%
  \institution{University of Bristol, UK}
}
\email{(eejlny,csxmh,ma17215)@bristol.ac.uk}

\author{Andr\'es Rodr\'{\i}guez, \\Rafael Asenjo, \\Angeles Navarro}
\affiliation{%
  \institution{Universidad de M\'alaga}
  \state{Spain} 
}
\email{(andres,asenjo,angeles)@ac.uma.es}

\author{Rub\'en Gran-Tejero, \\Dar\'io Su\'arez-Gracia}
\affiliation{%
  \institution{Universidad de Zaragoza}
  \state{Spain} 
}
\email{(rgran,dario)@unizar.es}

\renewcommand{\shortauthors}{Jose Nunez-Yanez et al.}


%
%
\begin{abstract}
In this paper, we introduce a software-defined framework that enables the parallel utilization of all the programmable processing resources available in heterogeneous system-on-chip (SoC) including FPGA-based hardware accelerators and programmable CPUs. Two platforms with different architectures are considered, and a single C/C++ source code is used in both of them for the CPU and FPGA resources.  Instead of simply using the hardware accelerator to offload a task from the CPU, we propose a scheduler that dynamically distributes the tasks among all the resources to fully exploit all computing devices while  minimizing load unbalance. The multi-architecture study compares an ARMV7 and ARMV8 implementation with different number and type of CPU cores and also different FPGA micro-architecture and size. We measure  that both platforms benefit from having the CPU cores assist FPGA execution at the same level of energy requirements.
\end{abstract}


\maketitle

\section{Introduction}
Heterogeneity is seen as a path forward for computers to deliver the energy and performance computing improvements needed over the next decade. In heterogeneous architectures, specialized hardware units accelerate complex tasks. A good example of this trend is the introduction of GPUs (Graphics Processing Units) for general purpose computing combined with multicore CPUs. FPGAs (Field Programmable Gate Arrays) are an alternative high performance technology that offer bit-level parallel computing in contrast with the word-level parallelism deployed in GPUs and CPUs.  In a typical configuration, the host CPU employs the FPGA accelerator to offload the work and then remains idle. In this research, we investigate a cooperative strategy applied to compute intensive applications in which both the CPU and FPGA perform the same task on different regions of the input data. The proposed scheduling algorithm dynamically distributes different chunks of the iteration space between CPU and a FPGA fabric integrated in the same die. The objective is to measure if simultaneous computing among these devices could be more favourable from an energy and/or performance points of view compared with offloading to the FPGA and the CPU idling. The FPGA and CPUs are programmed with the same C/C++ language using the SDSoC (Software Defined SoC) framework that enables very high productivity and simplifies the development of drivers to interface the processor and logic parts. As shown in Table \ref{tab:platformdetails}, we consider two platforms with different scales of compute power, one a low-cost platform with a dual-core ARMv7 CPU and another high-performance state-of-the-art platform with a quad-core ARMv8 CPU. Testing on both enables both the validation the approach and the comparison of their performance and energy characteristics. 

\



\begin{table}[hb]
\caption{Platform Specifications}
\label{tab:platformdetails}
\small
\begin{tabular}{p{2.2cm}p{2.8cm}p{2.8cm}}
\toprule
& \text{ZYNQ Z7020}        & \text{Zynq Ultrascale+ ZU9}   \\
\midrule
PL LUTs  & 53.2K & 274K \\
PL Flip-Flops  & 106.4K & 548K \\
PL Block RAMs  & 140 & 1824 \\
PL DSP Slices  & 220 & 2520 \\
Fabrication process  & 28 nm CMOS & 16 nm FinFET \\
PS, CPU type  & 32-bit dual Cortex A9 & 64-bit quad  
Cortex A53 \\
PS, CPU frequency & 600 MHz & 1.4 GHz \\
Nominal Voltage  & 1 Volt & 0.85 Volt \\
PL-PS interface  & Up to 4 64-bit HP ports & Up to 4 
128-bit HP ports\\
 & 1 64-bit ACP coherent port &  Up to 2 128-bit HPC coherent ports (no L2 allocation)\\
  &  & 1 128-bit ACP port (L2 allocation) \\
\bottomrule
\end{tabular}
\end{table}

\section{Background and related work}

The idea of balancing the workload among devices has been explored previously in the literature mainly around systems that combine GPUs and CPUs. For example, a study with desktop CPUs and GPUs has been done in~\cite{cgo14} where percentages of work to both devices are assigned before making a selection based on heuristics. With CPUs and GPUs, also energy aware decisions have been considered in~\cite{Dolbeau13}, which requires proprietary code. Another related work in the context of streaming applications~\cite{VilchesTPDS16} considers performance and energy when looking for the optimal mapping of pipeline stages to CPU and on-chip GPU. The possibility of using GPU+CPU and FPGA simultaneously and
collaboratively has also received attention in diverse application areas such as medical research~\cite{fpt12}. The hardware considered uses multiple devices connected through a common PCIe backbone, and the designers optimized how different parts of the application are mapped to each computing resource. This type of heterogeneous computing can be considered to connect devices vertically since the idea is to build a streaming pipeline with results moving processed data from one stage to the next. Data is captured and initially processed in the FPGA then moved with DMA engines to the CPU and GPU components. The
heterogeneous solution achieves a 273$\times$ speed-up over a multi-core CPU implementation. A study of the potential of FPGAs and GPUs to accelerate data center applications is done in~\cite{ispacs16}. The paper confirms that FPGA and GPU platforms can
provide compelling energy efficiency gains over general purpose processors, but it also indicates that the possible advantages of FPGAs over GPUs are unclear due to the similar performance per watt and the significant programming effort of FPGAs. In any case, it is important to note that the paper does not use high level languages to increase FPGA productivity as done in this work, and the power measurements for the FPGA are based on worst case tool estimations and not direct measurements. In this research, we explore a horizontal collaborative solution more closely related to the work done in~\cite{fpga10}. That work focuses on a multiple device solution similar to our work and demonstrates how the N-body simulation can be implemented in a heterogeneous solution in which both FPGA and GPU work together to compute the same algorithm kernel on different portions of particles. While our approach uses a dynamic scheduling algorithm to compute the optimal split, in~\cite{fpga10} the split is calculated manually with 2/3 of the workload given to FPGA and the rest to GPU; the collaborative implementation is 22.7$\times$ faster than the CPU only version.
In summary, we can conclude that the available literature has largely focused on advancing the programming models to make the use of FPGAs in heterogeneous systems more productive, comparing the performance of GPGPUs, FPGAs and CPUs for different types of applications in large scale clusters, and creating systems that manually choose the optimal device for each part of the application and move data among them. In contrast, in this paper we select a state-of-the-art high-level design flow based on C/C++ for single-chip heterogeneous CPU+FPGA and extend it to support simultaneous computing performing dynamic workload balancing. 


\section{Programming Environment}\label{sec:progenv}
\subsection{Programming Interface}\label{sec:interface}

\begin{figure*}[!htbp]
\includegraphics[width=\textwidth]{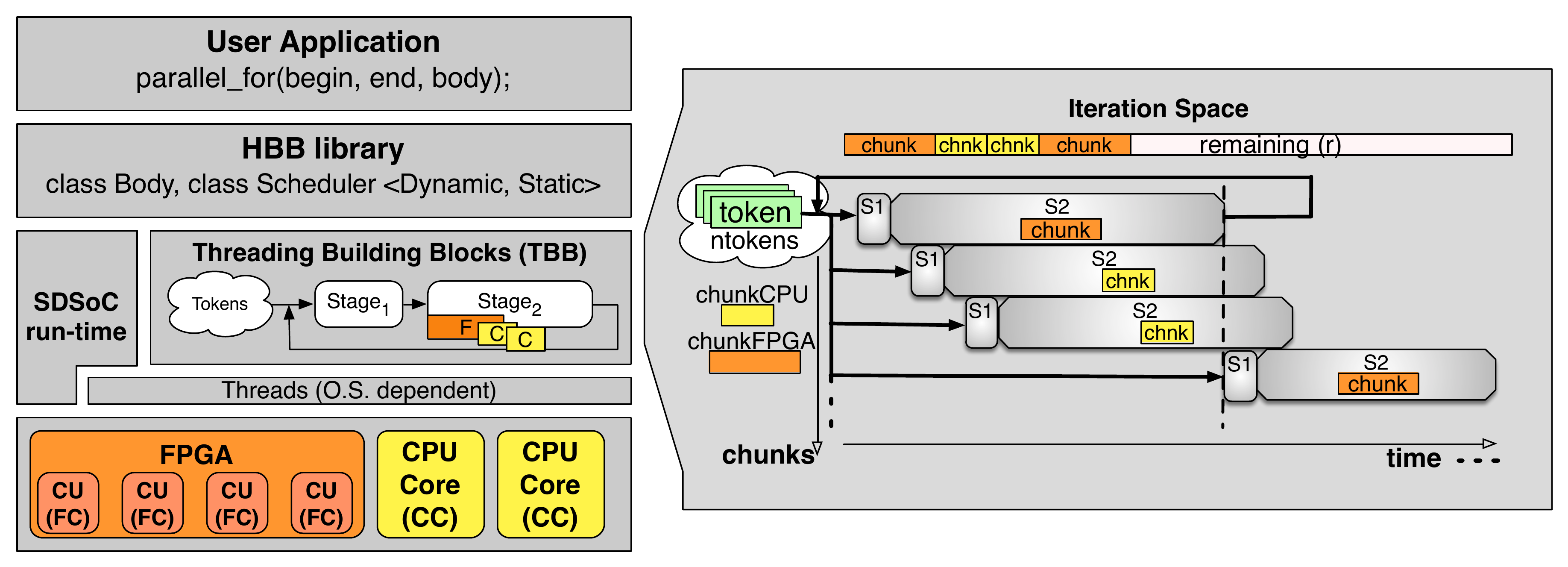}
\vspace*{-3mm}
\caption{Heterogeneous Scheduler}
\label{fig:scheduler}
\vspace*{-1mm}
\end{figure*}

This section introduces the proposed Heterogeneous Building Blocks (\textbf {HBB})
library API. It is a C++ template library that takes advantage of heterogeneous
processors and facilitates his usage and configuration. HBB aims to make easier
the programming for heterogeneous processors by automatically partitioning and
scheduling the workload among the CPU cores, and the accelerator. It builds on top of the
SDS (Xilinx SDSoC library) and TBB\cite{TBB} libraries, and it offers a \texttt{parallel\_for()} function
template to run on heterogeneous CPU+FPGA systems. In
Fig.~\ref{fig:scheduler} we depict an MPSoC with an integrated FPGA and two CPU cores (CC), as the low-end platform used in the experimental evaluation. The FPGA itself can contain a number of FPGA compute units (FC) depending on resource availability and accelerator configuration. 

The left part of Fig.~\ref{fig:scheduler} shows the software stack that
supports the user application. Our library (HBB) offers an abstraction layer
that hides the initialization and management details of TBB and SDS
constructs, thus the user can focus
on his own application instead of dealing with thread management and
synchronization. The library takes care of splitting the iteration space in chunks of iterations and process each chunk on a CPU core (CC) or a FPGA compute unit (FC). The size of the chunks that are offloaded to the FC is constant an provided by the user so that it is big enough to fully utilize the FC, but small enough to foster work sharing and load balance among the CCs and FCs. The size of the chunks processed on the CCs is adaptively computed by our heterogeneous scheduler as explained in Section~\ref{sec:partitioner}. The right part of Fig.~\ref{fig:scheduler} shows that the
internal engine that manages the \texttt{parallel\_for()} function is a two-stage
pipeline, Stage$_1$(S1) and Stage$_2$(S2), implemented with the TBB pipeline template. At the top of
this part we can see the iteration space with the chunks that have already been
assigned to a processing resource (in orange for the FPGA and yellow for the two CPU
cores) and the remaining iterations with the iterations that have not been
assigned yet (in white). The right part of the figure shows an execution of the
pipeline with 3 \texttt{tokens}. The tokens represent the
number of chunks of iterations that are processed in parallel.
The time required
for the computation of each processed chunk on a FC or on a CC is recorded. This
time is used to update the relative speed of the FC w.r.t.
a CC, that we call $f$. Factor $f$ will be required to adaptively
adjust the size of the next chunk assigned to a CC as we will see in Section~\ref{sec:partitioner}.

\begin{figure}[hbt]
\centering
\begin{minipage}{0.9\linewidth}
\begin{lstlisting}[style=C]
#include "hbb.h"

int main(int argc, char* argv[]){
  Body body; (*@\label{list:body}@*)
  Params p; (*@\label{list:params}@*)
  InitParams (argc, argv, &p);(*@\label{list:initparams}@*)
  // Instantiate task scheduler
  Dynamic * hs = Dynamic::getInstance(&p);(*@\label{list:initd}@*)
  ...
  hs->parallel_for(begin, end, body);(*@\label{list:for}@*)
  ...
}

\end{lstlisting}
\end{minipage}
\vspace*{-4mm}
\caption{Using the \texttt{parallel\_for()} function template}
\label{fig:parallelforcode}
\vspace*{-1mm}
\end{figure}

Fig.~\ref{fig:parallelforcode} shows a main function with all the required
component initialization to make the \texttt{parallel\_for()}
function template works. This is the main component of the
HBB library and it is made available by including the hbb.h
header file. The user has to create a \texttt{Body} instance (line~\ref{list:body}) that will later be passed to the \texttt{parallel\_for()} function. Program
arguments, like the number of threads and scheduler configuration
can be read from the command-line, as can be seen in
line~\ref{list:initparams}. The benchmarks that we evaluate accept at
least three command-line arguments: \texttt{<num\_cpu\_t>},
\texttt{<num\_fpga\_t>} and \texttt{<fpga\_chunksize>}. The first one sets
the number of CPU tokens, which translates into how many
CPU cores will be processing chunks of the iteration space.
The second one can be set just to 0 or 1 to disable or not
the FPGA as an additional computing resource. The last argument,
\texttt{<fpga\_chunksize>} set the number of iterations that will contain the chunks offloaded to the FPGA.

\begin{figure}[hbt]
\centering
\begin{minipage}{0.9\linewidth}
\begin{lstlisting}[style=C]
class Body{(*@\label{list:bodyclassdef}@*)

public:
  void operatorCPU(int begin, int end) { (*@\label{list:operatorCPUstart}@*)
     for(i=begin; i!=end; i++){
        c[i] = a[i] * b[i]; }
  }(*@\label{list:operatorCPUend}@*)

  void operatorFPGA() (int begin, int end){(*@\label{list:operatorFPGAstart}@*)
    mmult((float*)array_a,(float*)array_b,(float*)array_c, begin, end, scalar, status, enable);
  }(*@\label{list:operatorFPGAend}@*)
};
...
\end{lstlisting}
\end{minipage}
\vspace*{-3mm}
\caption{Definition of Class Body}
\label{fig:bodyclass}
\vspace*{-1mm}
\end{figure}


Before using the \texttt{parallel\_for()} function, the user must implement a
\texttt{Body} class in order to define the body of
the parallel loop, as we see in Fig.~\ref{fig:bodyclass}. This class must implement two methods: one that defines the
code that each CPU core has to execute for an arbitrary chunk of iterations, and
the same for the FPGA device. The \texttt{operatorCPU()} method
(lines \ref{list:operatorCPUstart}-\ref{list:operatorCPUend} in
Fig.~\ref{fig:bodyclass}) defines the CPU code of the kernel, and the
\texttt{operatorFPGA()} method (lines
\ref{list:operatorFPGAstart}-\ref{list:operatorFPGAend}) calls a hardware function that has been already implemented in the FPGA using the SDSoC development flow. SDSoC automatically manages the data movement from global memory to the FPGA and back.

\subsection{Scheduling strategies}\label{sec:partitioner}

This section covers the computation of the chunk size that will be
executed by the CPU cores and the FPGA. We implement different scheduling policies, but in this work we focus in the dynamic scheduling strategy. 


When the dynamic scheduling is selected (see line~\ref{list:initd} in Fig.~\ref{fig:parallelforcode}), then the argument \texttt{<fpga\_chunksize>} sets the FPGA chunk size, $S_f$, 
whereas the CPU chunk size is automatically computed by a heuristic described in~\cite{TRNavarro} and briefly summarized next. This heterogeneous dynamic scheduler is a combination of the OpenMP dynamic scheduler \cite{openmp} for the FPGA chunks and the OpenMP guided scheduler for the CPU chunks. 
Assuming that $n$ is the number of iterations of the \texttt{parallel\_for()}, $nCores$ the number of CPU cores, and $r$ the number of remaining
iterations (initially $r = n$), then the computation of the CPU chunk,
$S_c$, follows the next expression:
\vspace*{-1mm}
\begin{equation*}
\footnotesize
\centering
S_c=min\Bigl(\frac{S_f}{f},\frac{r}{f+nCores}\Bigr)
\label{eq:line6}
\end{equation*}

\normalsize
\noindent where $f$ represents how much faster the FPGA is w.r.t. a CPU core, and it is
recomputed each time a chunk is processed, as explained in Section~\ref{sec:interface}. In other words, $S_c$ is
either $(S_f/f)$ (the number of iterations that a CPU core must perform to
consume the same time as the FPGA) when the number of remaining iterations, $r$,
is sufficiently high, or $r/(f+nCores)$ (a \textit{guided self-scheduling
strategy}~\cite{Rudolph:ICS89}), when there are few remaining iterations, this
is when $r/(f+nCores) < S_f/f$. 

\section{Benchmark development}

This preliminary evaluation is based on a well-known benchmark: GEMM (General Matrix Multiplication). The benchmark is written in C/C++ for both FPGA and CPU targets, and the FPGA functions are compiled using the high-level synthesis tools that are part of the SDSoC framework. 

\begin{figure*}[!htbp]
\includegraphics[width=.8\textwidth]{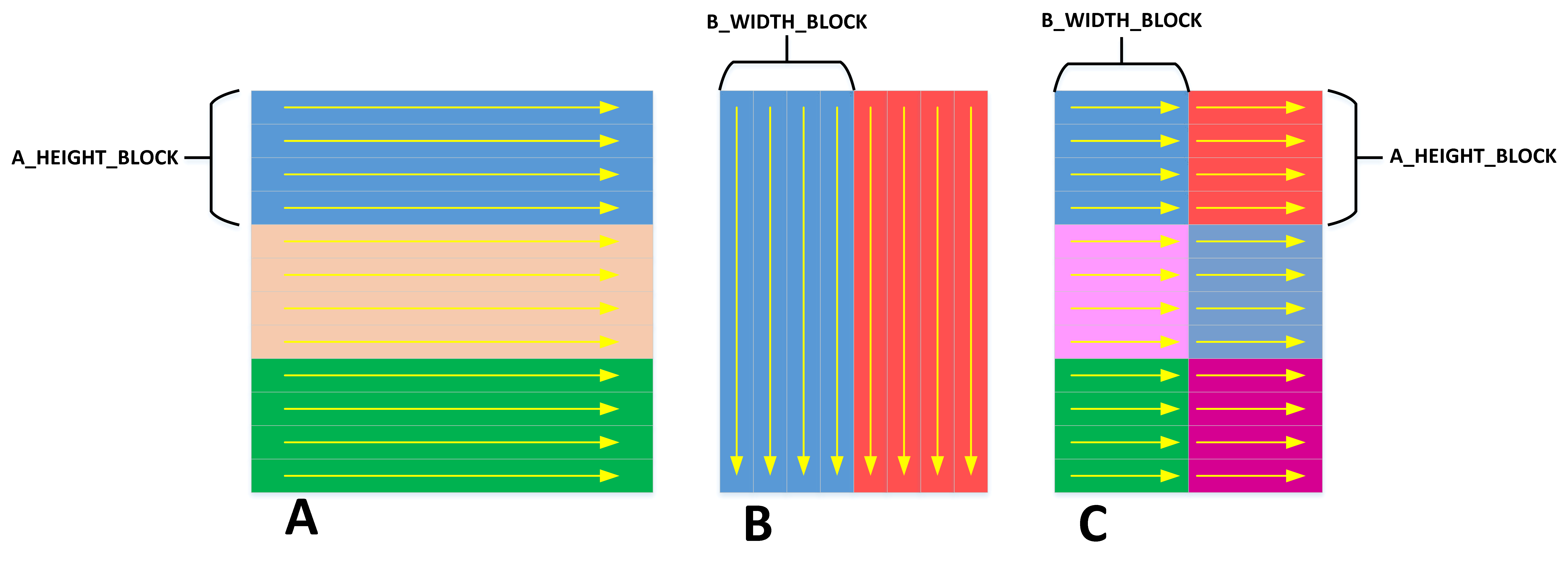}
\caption{Matrix multiplication tiling}
\label{fig:mmt}
\end{figure*}

The algorithm is based on a tiling strategy depicted in Figure~\ref{fig:mmt} in which the matrix blocks are shown with different colors. A and B are the input matrices and C is the output matrix. For example multiplying the green block of A with the red block of B will generate the purple block of C. The matrix size used in the main experiment of 1M elements cannot be buffered completely in FPGA memory so the tiling strategy becomes necessary. Matrix B cannot be declared as having sequential access in SDSoC because the blocks inside matrix B are not accessed in a sequential manner and for this reason DMA options are not possible. Matrix A is accessed sequentially but it is read multiple times. For that reason it cannot be declared as having a sequential access either. The multiple reads of the same matrix during a single multiplication will not work with the DMA correctly. Notice that sequential access is needed to use a DMA solution based on either SDSoC scatter\_gather or simple\_dma. Both use virtual addresses that must be sequential although scatter\_gather allows physical addresses that are non\-sequential. Since using SDSoC DMAs is not possible in this benchmark the interfaces are based on AXIMM (AXI Memory Master) that can also obtain high performance using the long burst modes available in AXI. 

Table \ref{tab:gemmhw} shows the results of using the same source code for both devices while varying the number matrix B columns that are buffered inside the FPGA (32 in case of the Zynq and 128 in case of the Zynq Ultra). As the number of buffered elements increases it is possible to extract more parallelism. The available internal memory in the Zynq device limits this value to 32 but in the case of the Zynq Ultra the 128 value is due to a tool issue that fails to perform synthesis with larger values than 128. The Zynq device can only accommodate one single FPGA compute unit while the Zynq Ultra supports the deployment of 4 compute units, working in parallel. To enable cache coherence the ACP port is used in the Zynq device and the HPC ports are used in the Zynq Ultra device. Cache coherence is important when the application requires CPU and FPGA cores have access to the same data to guarantee correctness and to avoid explicit software coherency. 

\begin{table}[htb]
\caption{GEMM hardware resources}
\label{tab:gemmhw}
\begin{tabular}{lcccc}
\toprule
    & \multicolumn{2}{c}{Zynq Ultra}   & \multicolumn{2}{c}{Zynq} \\ 
    & \text{available} & \text{used} / \% & \text{available} & \text{used} / \% \\
\midrule
LUTs (K)  & 274 & 87.8 / 32.0 & 53.2 & 18.1 / 34.0\\
Flip-Flops (K) & 548.1 & 162.6 / 29.7 & 106.4 & 27.3 / 25.7\\
Block RAMs  & 1824 & 1048 / 57.5 & 140 & 79 / 56.4 \\
DSP Slices  & 2520 & 640 / 25.4 & 220 & 160 / 72.7\\
\bottomrule
\end{tabular}
\end{table}

\section{Heterogeneous computing evaluation}

The evaluation of the GEMM benchmark is performed on a ZC702 board equipped with a Zynq 7020 device and the ZCU102 board equipped with a Zynq Ultrascale Z9 device. These board contains a PMBUS (Power Manager BUS) power control and monitoring system that enables the reading of power and current values using the ARM CPUs. For the power measurements the values of power corresponding to the processing system (CPU cores), programmable logic (FPGA) have been added together. For the energy computation we multiply this value for the execution time of the benchmark. 

\begin{figure*}[!htbp]
\centering
\begin{subfigure}{.5\textwidth}
  \centering
  \includegraphics[width=.99\columnwidth,height=4.5cm]{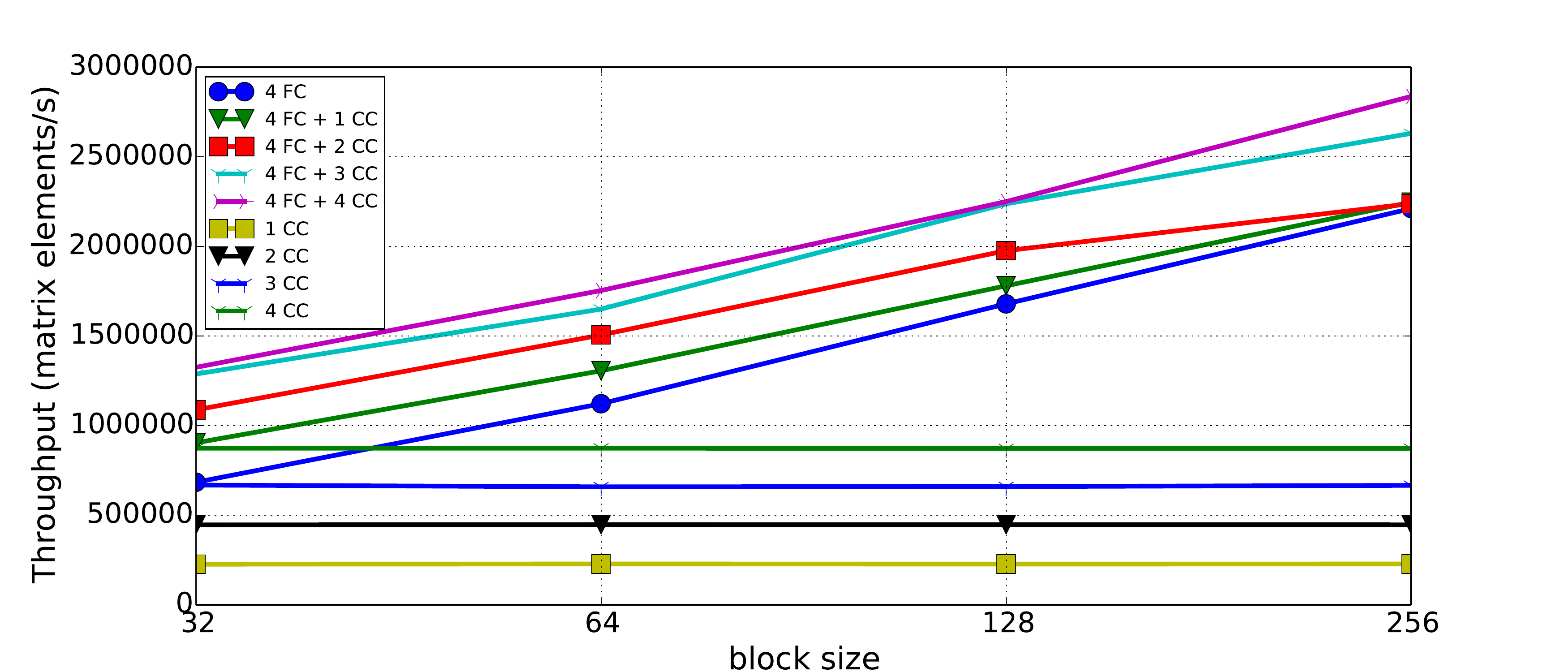}
  \caption{GEMM ZYNQ Ultrascale}
  \label{fig:perf_aes}
\end{subfigure}%
\begin{subfigure}{.5\textwidth}
  \centering
  \includegraphics[width=.99\columnwidth]{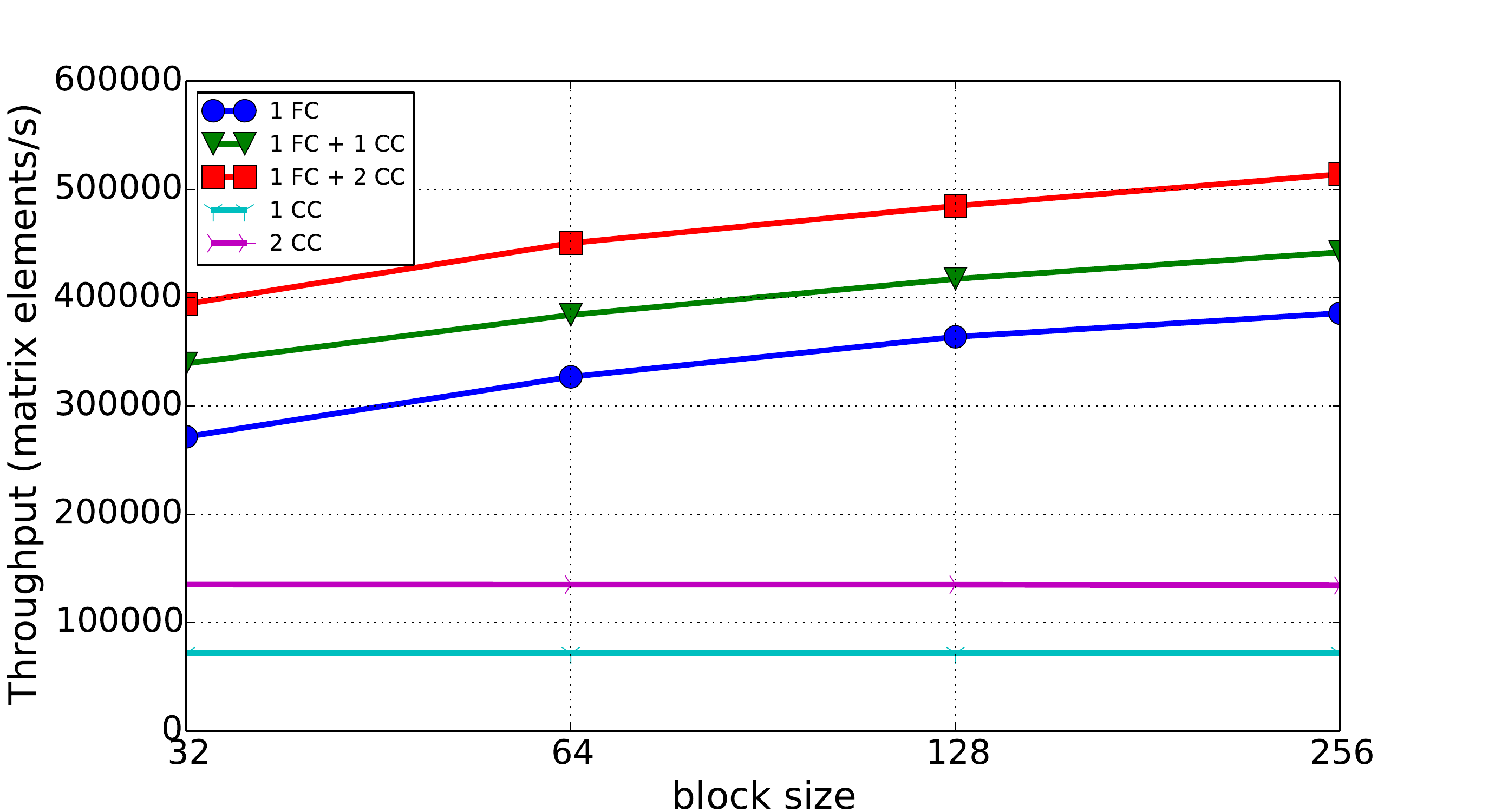}
  \caption{GEMM ZYNQ}
  \label{fig:perf_gemm}
\end{subfigure}
\caption{Benchmarks performance analysis}
\label{fig:fig_performance}
\end{figure*}

\begin{figure*}[!htbp]
\centering
\begin{subfigure}{.5\textwidth}
  \centering
  \includegraphics[width=.99\columnwidth]{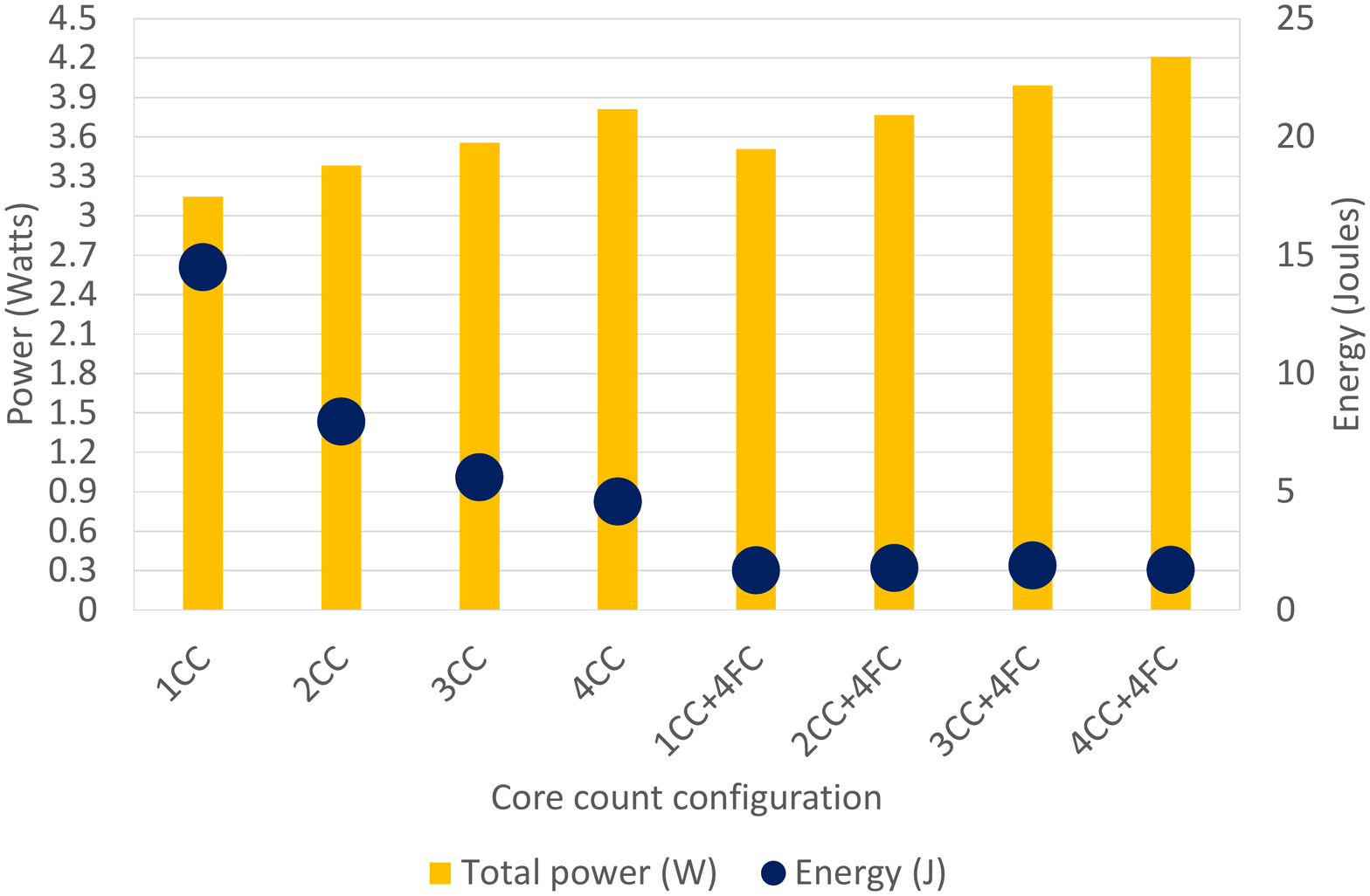}
  \caption{GEMM ZYNQ Ultrascale}
  \label{fig:power_aes}
\end{subfigure}%
\begin{subfigure}{.5\textwidth}
  \centering
  \includegraphics[width=.99\columnwidth]{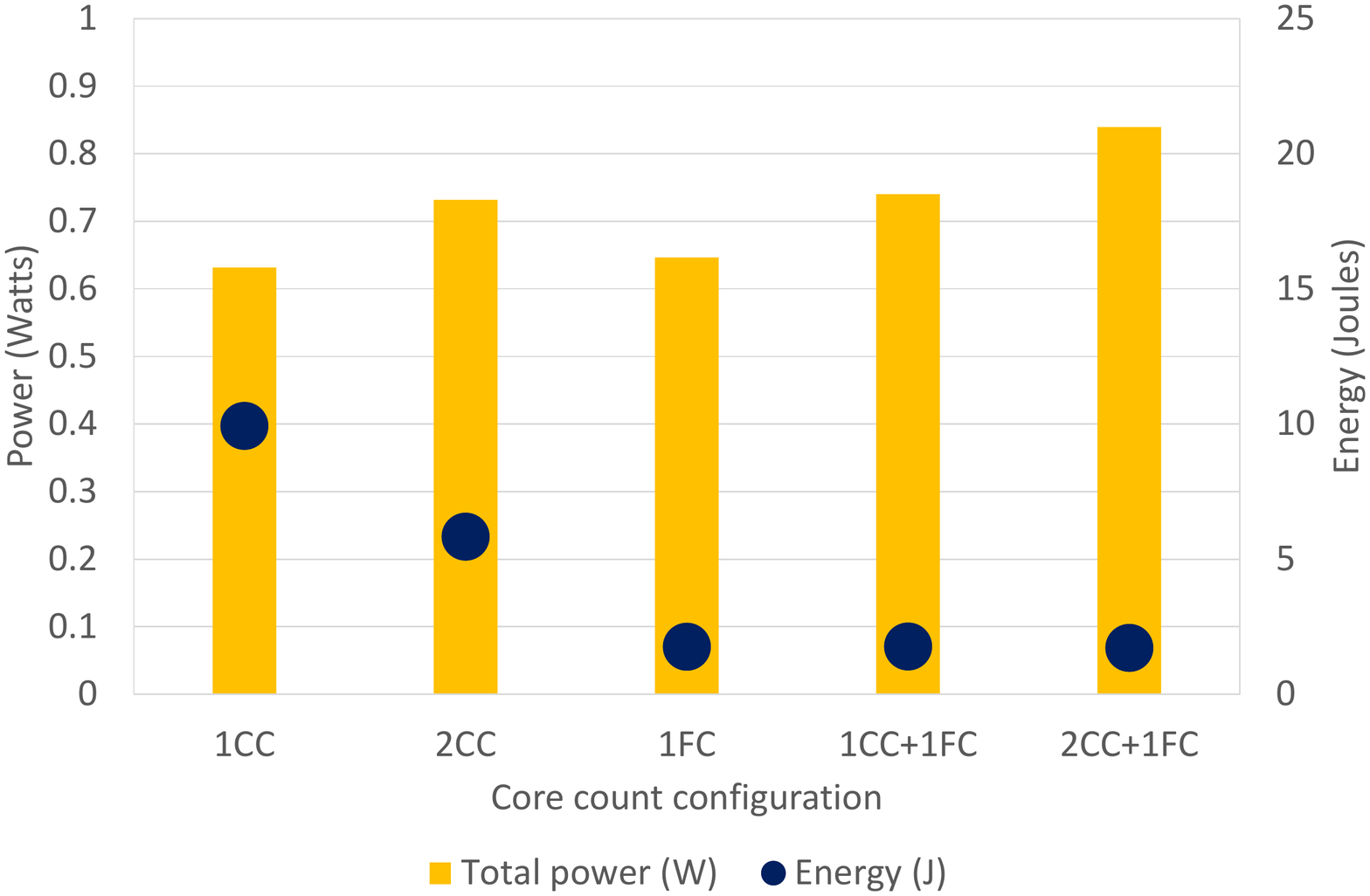}
  \caption{GEMM ZYNQ}
  \label{fig:power_gemm}
\end{subfigure}
\vspace*{-2mm}
\caption{Benchmarks power and energy}
\label{fig:fig_power}
\end{figure*}

Figs.~\ref{fig:fig_performance} and~\ref{fig:fig_power} show performance, power and energy consumption when we explore different chunk sizes for the FPGA (X axis) in our dynamic scheduling strategy with a fixed matrix size of 1M elements. Note that the CPU chunk sizes are determined adaptively, as explained in Section~\ref{sec:partitioner}. Different configurations are evaluated and the number of active CPU cores (CC) and FPGA compute units (FC) ranges from 0 to 4. 

Fig.~\ref{fig:fig_performance} shows the performance evaluation of the GEMM benchmark. The heterogeneous configurations are the fastest for both Zynq and Zynq Ultrascale. Overall the Zynq Ultrascale configuration is up to 6.5 times faster than the Zynq device and the highest performance is achieved with 4 CPU cores and the 4 FPGA cores in parallel. 

Fig.~\ref{fig:fig_power} compares the energy and power results for both systems. The Zynq Ultrascale device highest power usage is 4.2 Watts while Zynq uses 0.8 Watts. This means that power usage is 5.25 higher in the Zynq Ultrascale device and this increase in power means that the energy values are comparable in both devices. We believe that as the the Zynq Ultrascale compiler improves and larger configurations are possible, the speed-up factor should increase and differences in energy efficiency should be more noticeable.  Initial results with a matrix size of 16M elements show that the performance of the Zynq platform drops from 500K to 50K matrix elements per second while the Zynq Ultra reaches 400K matrix elements per second which is 8x higher.   



\section{Conclusion}

This paper has presented initial results of a dynamic scheduler that shares work on FPGA+CPU system-on-chips improving performance at the same level of energy consumption. Two hybrid CPU+FPGA SoCs with different CPU and FPGA microarchitectures and resources with the same single-source programming model are compared in terms of performance, power and energy. The experiments show that a noticeable performance gain can be achieved in  both platforms with heterogeneous computing. Heterogeneous configurations that allow the CPU cores collaborate with the FPGA reduce execution times from 25\% to 50\%. If the objective is to minimize energy, then the heterogeneous versions tend to be energy neutral since the additional power required by the CPU cores is compensated by the reduction in execution time.  The more powerful Ultrascale platform is significantly faster in terms of performance but the additional CPU and FPGA static and dynamic power suggests that it will be necessary to achieve performance speed-ups higher than one order of magnitude to observe meaningful energy savings. Future work includes the generalization of the methodology to other benchmarks, larger workloads and exploring the additional PL-PS interfaces available in the system.


\section*{Acknowledgment}
This work was partially supported by Xilinx, the Spanish projects TIN 2013-42253-P,  P11-TIC-08144, TIN2013-46957-C2-1-P, TIN2016-76635-C2-1-R, gaZ: T48 research group and UK EPSRC with the ENPOWER (EP/L00321X/1) and the ENEAC (EP/N002539/1) projects.
\bibliographystyle{plain} 
\vspace*{-0.2cm}
\bibliography{paper}
\end{document}